\documentclass[a4paper]{article}
\usepackage{xcolor}
\usepackage[utf8]{inputenc}
\usepackage[T1]{polski}
\usepackage{graphicx}\graphicspath{{img/}}
\usepackage{lmodern,fancyhdr,secdot,float,amsmath,amsfonts,amsthm,amssymb}
\usepackage{booktabs,tabularx}
\usepackage[margin=2cm]{geometry}
\usepackage{multicol}
\usepackage[hidelinks,unicode]{hyperref}
\usepackage{titlesec}
\titleformat{\section}{\centering\normalsize\bf}{\thesection.}{.5em}{\MakeUppercase}
\titleformat*{\subsection}{\bf\normalsize\selectfont}
\titleformat*{\subsubsection}{\bf\normalsize\selectfont}
\newcommand{\titlePL}[1]{\large\textbf{ #1}}
\newcommand{\titleEN}[1]{\normalsize #1}
\newcommand{\keywordsPL}[1]{\small\textbf{Słowa kluczowe:} #1}
\newcommand{\keywordsEN}[1]{\small\textbf{Keywords:} #1}
\newcommand{\abstractPL}[1]{\small\textbf{Streszczenie:} #1}
\newcommand{\abstractEN}[1]{\small\textbf{Abstract:} #1}

\usepackage{algorithm}
\usepackage{algorithmic}

\definecolor{logo_color}{RGB}{40, 69, 166}

\usepackage[numbers]{natbib}
\begin{document}\thispagestyle{empty}\pagestyle{fancy}
\begin{minipage}[t]{0.5\textwidth}\vspace{0pt}%
\includegraphics[scale=0.9]{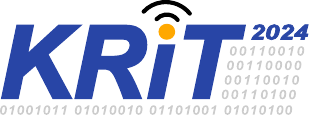}
\end{minipage}
\begin{minipage}[t]{0.45\textwidth}\vspace{12pt}%
\centering
\color{logo_color} KONFERENCJA RADIOKOMUNIKACJI\\ I TELEINFORMATYKI\\ KRiT 2024
\end{minipage}

\vspace{1cm}

\begin{center}
\titlePL{Wykorzystanie powietrznych przełączalnych inteligentnych powierzchni do komunikacji międzypojazdowej}

\titleEN{Deploying an Aerial Reconfigurable Intelligent Surface for
Vehicle-to-Vehicle Communications}\medskip

Salim Janji

\medskip

\begin{minipage}[t]{0.8\textwidth}
\centering
\small Institute of Radiocommunications, Poznan\\ \href{mailto:email}{salim.janji@put.poznan.pl} \\
\end{minipage}
\medskip
\end{center}

\medskip

\begin{multicols}{2}
\noindent
\abstractPL{
Niniejszy artykuł omawia wdrożenie drona wyposażonego w rekonfigurowalną inteligentną powierzchnię (Reconfigurable Intelligent Surface, RIS) celem stworzenia dronowej stacji przekaźnikowej (Drone Relay Station, DRS), aby zwiększyć łączność między pojazdami (Vehicle-To-Vehicle, V2V) na ziemi. Trajektoria DRS jest optymalizowana w taki sposób, aby możliwie najszybciej osiągnąć położenie maksymalizujące przepustowość. Dodatkowo uwzględniana jest obecność węzła zakłócającego, a rozwiązanie analityczne jest wyprowadzone w celu określenia optymalnej orientacji DRS w każdym kroku czasowym, minimalizując zakłócenia dla odbiornika. Wyniki symulacji potwierdzają skuteczność proponowanego modelu.}
\medskip

\noindent
\abstractEN{
This paper addresses the deployment of a drone equipped with a reconfigurable intelligent surface (RIS), creating a drone relay station (DRS) to enhance the connectivity of vehicle-to-vehicle (V2V) pairs on the ground. The trajectory of the DRS is optimized to quickly reach the best location for maximizing throughput. Additionally, the presence of an interfering node is considered, and an analytical solution is derived to determine the optimal orientation of the DRS at each time step, minimizing interference to the receiver. Simulation results confirm the effectiveness of the proposed framework.
\footnote{Copyright © 2024 SIGMA-NOT. Personal use is permitted. For any other purposes, permission must be obtained from the SIGMA-NOT by emailing sekretariat@sigma-not.pl. This is the author’s version of an article that has been published in the journal entitled \textit{Telecommunication Review -- Telecommunication News} (PL: \textit{Przegląd Telekomunikacyjny -- Wiadomości Telekomunikacyjne}) by the SIGMA-NOT. Changes were made to this version by the publisher before publication, the final version of the record is available at: https://dx.doi.org/
10.15199/59.2024.4.80. To cite the paper use: S. Janji, “Deploying an Aerial Reconfigurable Intelligent Surface for Vehicle-to-Vehicle Communications" (PL: "Wykorzystanie powietrznych przełączalnych inteligentnych powierzchni do komunikacji międzypojazdowej"), \textit{Telecommunication Review -- Telecommunication News} (PL: \textit{Przegląd Telekomunikacyjny -- Wiadomości Telekomunikacyjne}), 2024, no. 4, pp.~360--363, doi: 10.15199/59.2024.4.80 or visit https://sigma-not.pl/publikacja-150619-2024-4.html.}}
\medskip

\noindent
\keywordsPL{bezzałogowy statek powietrzny, rekonfigurowalna inteligentna powierzchnia, niezawodność komunikacji, zakłócenia}
\medskip

\renewcommand{\figurename}{Figure}

\noindent
\keywordsEN{UAV, RIS, communication reliability, interference}

\section{Introduction}

The emergence of intelligent transportation systems, including autonomous vehicles, has propelled Vehicle-to-Everything (V2X) technology as a key enhancer of road safety, traffic efficiency, and passenger connectivity. V2X incorporates various communication modes such as Vehicle-to-Vehicle (V2V), Vehicle-to-Infrastructure (V2I), Vehicle-to-Pedestrian (V2P), and Vehicle-to-Network (V2N), all demanding robust, low-latency, and high-throughput communication capabilities for applications like collision avoidance, traffic management, and autonomous driving \cite{survey_1, survey_2}.

In areas with connected vehicles, efficient V2V networks are essential, particularly when spectrum availability is limited or vehicle traffic is high. Obstacles blocking the line-of-sight (LOS) path between vehicles pose significant challenges by reducing signal strength. The deployment of drones, or unmanned aerial vehicles (UAVs), has been explored as a solution to support V2V communication by acting as relay nodes \cite{uav_v2v, uav_v2v_2}.

Additionally, reconfigurable intelligent surfaces (RISs) are becoming a popular tool for network enhancement by modulating signal strength, coverage, and energy efficiency \cite{ris_tutorial}. RISs, which are embedded with elements that control the amplitude and phase of electromagnetic waves, can direct signals to specific areas, improving desired signal reception and reducing interference. The integration of RISs on drones to support ground connectivity has recently been investigated, offering a potential solution to bandwidth limitations and signal blockages \cite{uav_ris, uav_ris_2, uav_ris_3, uav_ris_own}.

\begin{figure}[H]
\centering
\includegraphics[width=0.95\linewidth,trim={0cm, 0cm, 0cm, 0cm},clip]{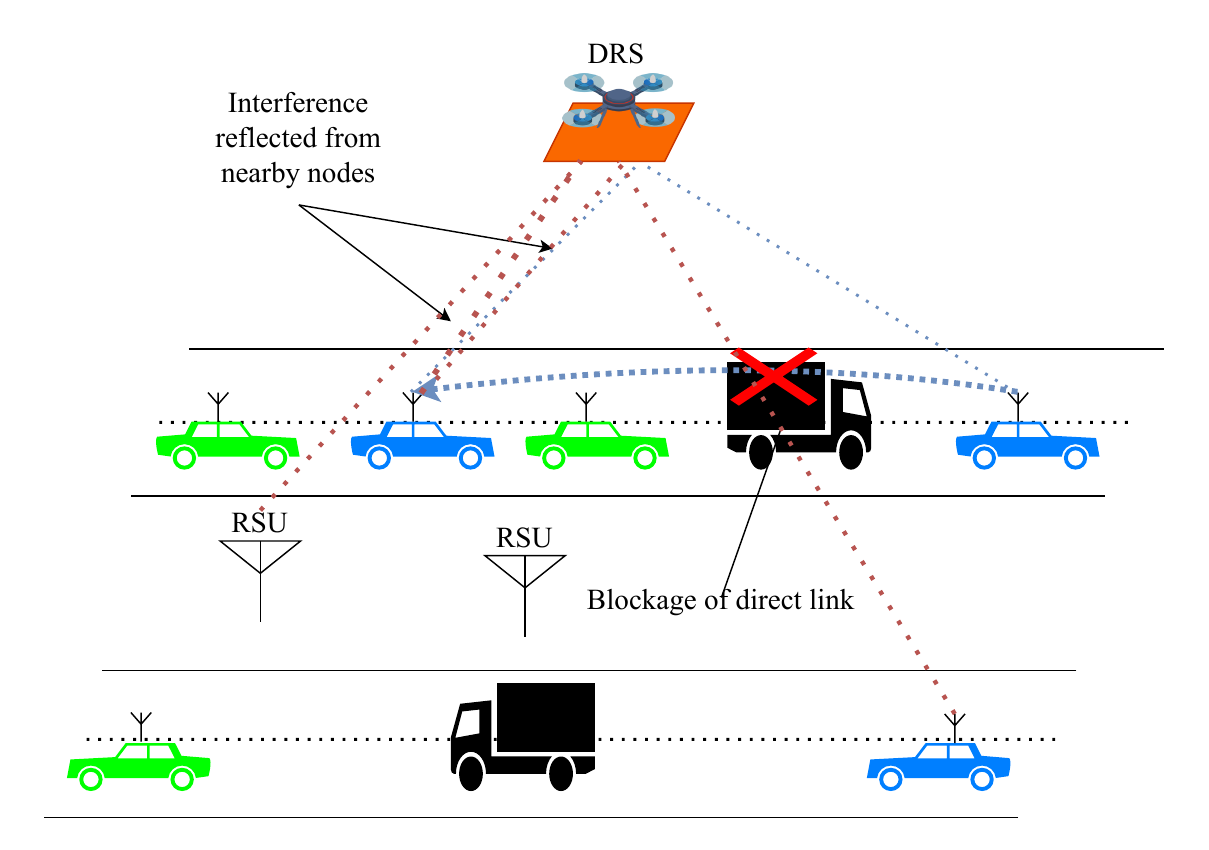}
\caption{Depiction of the examined scenario where a V2V pair is engaged in communication with potential obstacles impeding direct connections between vehicles. A DRS is deployed to enhance connectivity, yet interference reflected by the DRS from other nodes must also be accounted for.}
\label{fig:scenario}
\end{figure}

This paper examines the deployment of a drone relay station (DRS) equipped with an on-board RIS to aid V2V communications on the ground. The previous study in \cite{uav_ris_own} addressed this problem for a single pair of vehicles without considering interference from other nodes. In contrast, this work considers the interference from a nearby transmitting vehicle or road-side unit (RSU) and includes the optimization of the RIS orientation to minimize interference to the receiver of the serviced pair. The scenario is illustrated in Fig.~\ref{fig:scenario}.

The structure of the paper is as follows: The next section introduces the system model and discusses the channel model. Section~3 presents the problem formulation, followed by Section~4, which details the proposed solutions for trajectory planning and orientation control. An analysis of the simulation results follows. The paper concludes with a summary of the findings.

\section{System model}
\label{sec:system_model}
This study explores a V2V communication scenario with vehicles traveling in two lanes, as depicted in Fig.~\ref{fig:scenario}. It is assumed that $K=1$ pair of vehicles communicate using the same frequency band. The 3D locations of the vehicles at time $t$ are given by $(\mathbf{L}_{1}(t), \mathbf{L}_{2}(t))$. The position of each vehicle or RSU is represented as $\mathbf{L}_{j}(t) = (x_j(t), y_j(t), h_j(t))$, where $x_j(t)$, $y_j(t)$, and $h_j(t)$ denote the lateral, longitudinal, and vertical positions, respectively. Similarly, an interfering node, which can be either an RSU or another vehicle, is identified with its location, $\boldsymbol{L}_{\text{I}}(t)$. All vehicles maintain a constant velocity $v_j$, and their antennas are positioned between 1.5 and 2 meters high.

Direct V2V links may not provide sufficient reliability or throughput. Therefore, the use of a UAV equipped with a nearly-passive RIS, referred to as a DRS, is considered to enhance signal reflection. The UAV's position is defined by $\mathbf{L}_{D}(t) = (x_D(t), y_D(t), h_D(t))$, and its velocity $v_D$ is greater than or equal to any vehicle's velocity $v_i$. The RIS's reflection capabilities depend on its orientation matrix $\mathbf{O}(t)$, a $3 \times 3$ matrix determining the alignment of the RIS.


To model RIS-aided transmission, the methodologies from \cite{Tang2021} for far-field or near-field beamforming, based on the relationship between transmitter-to-RIS distance $d_{1}$, RIS-to-receiver distance $d_{2}$, and the Faunhofer distance $d_{Fr} = \frac{2D^2}{\lambda}$, are employed. Given the setup, all transmissions use far-field beamforming. The effective path-loss in this scenario is influenced by the alignment of the transmission path with the RIS orientation, accounted for by the azimuth and elevation angles of both transmitter and receiver. The applicable path-loss formula is presented as:

\begin{equation}
\begin{split}
    PL_{f-f} &=  \frac{64 \pi^3 d_{1}^2 d_{2}^2 }{G_t G_r G M^2 N^2 d_{x} d_{y} \lambda^2 F(\theta_t), F(\theta_r) A^2 \left\vert\Psi\right\vert^2}\\
    \Psi = & \frac{sinc(\frac{M\pi}{\lambda}(\sin{\theta_t}\cos{\varphi_t}+\sin{\theta_r}\cos{\varphi_r})d_x)}{sinc(\frac{\pi}{\lambda}(\sin{\theta_t}\cos{\varphi_t}+\sin{\theta_r}\cos{\varphi_r})d_x)} \\
    &\times \frac{sinc(\frac{N\pi}{\lambda}(\sin{\theta_t}\sin{\varphi_t}+\sin{\theta_r}\sin{\varphi_r})d_y)}{sinc(\frac{\pi}{\lambda}(\sin{\theta_t}\sin{\varphi_t}+\sin{\theta_r}\sin{\varphi_r})d_y)},
\end{split}
\label{eq:ris_ff_ploss}
\end{equation}
where $G_t$, $G_r$, and $G$ denote the antenna gains of the transmitter, receiver, and RIS, respectively; $M$ and $N$ represent the rows and columns of RNA elements; $d_x$ and $d_y$ are the element spacings; $F(\theta_t)$ and $F(\theta_r)$ are the normalized radiation patterns directed towards the transmitter and receiver. All RIS elements share a uniform radiation pattern, defined as:
\begin{equation}
    F(\theta) = \begin{cases}  
    \cos^3{\theta} & \theta \in \left[0, \frac{\pi}{2}\right]\\
    0 & \theta \in \left(\frac{\pi}{2}, \pi\right]
\end{cases}
\label{eq:ris_radiation}
\end{equation}

The throughput for each pair is calculated using the modified Shannon formula, incorporating the signal-to-interference-plus-noise ratio (SINR), bandwidth, and path-loss:
\begin{equation}
\begin{split}\label{eqn:rate}
R = \eta B_{\text{eff}}\log_2 (1+SINR)
\end{split}
\end{equation}
\begin{equation}
SINR = \frac{P_t}{PL \sigma_{N}^2 + \frac{P_t}{PL_I}}
\end{equation}
where $\eta$ and $B_{\text{eff}}$ denote the link effectiveness and effective bandwidth, respectively, $PL$ represents the path-loss between the communicating pair of vehicles, $PL_I$ is the path loss from the interfering node to the receiver (the receiver is identified by $L_2$), and $P_t$ is the transmission power.

\section{Problem formulation}
\label{sec:problem}

The instantaneous positions of a V2V communication pair, represented as ($\boldsymbol{L}_{1}(t)$, $\boldsymbol{L}_{2}(t)$), along with the position of the DRS, $\boldsymbol{L}_{\text{D}}(t)$, the position of the interfering node, $\boldsymbol{L}_{\text{I}}(t)$, and the 3x3 orientation matrix $\boldsymbol{O}(t)$, are considered. The task is to design the future trajectory of the DRS, specifically determining $\boldsymbol{L}_{\text{D}}(t + nT_s)$ for $n \in \{1, 2, \ldots, N\}$, where $T_s$ is the planning time step, and $N$ is the number of steps, determined by the duration of the V2V pair's transmission or their presence within the designated highway segment, which is confined by $(x_{\min}, x_{\max})$ on the $x$-axis and $(y_{\min}, y_{\max})$ on the $y$-axis. Additionally, the DRS is capable of modifying the orientation of the RIS along the \textit{xy}-plane, denoted by $\boldsymbol{O}(t)$, at a rotation rate of $\Gamma_D$ $[\frac{rad}{s}]$. It is essential to establish not only the trajectory but also the orientation $\boldsymbol{O}^{(n)}$ at each time step. For clarity and consistency throughout this paper, the time index is represented as a superscript in all subsequent references. Uniform direction and speed $[\frac{m}{s}]$ within each time step are assumed, constrained by the maximum speed $v_D$.

The objective is to maximize the achieved rate through minimizing the path loss of the virtual LOS link, established through the RIS, while also minimizing the reflected interference from a nearby node. The transmission rate $R^{(n)} $ is determined using \eqref{eqn:rate} based on the vehicles' locations and the DRS's position and orientation. The optimization problem is thus formulated as:

\begin{subequations}\label{eqn:optimization0}
\begin{align}
    &\;\;\;\;\;\;\;\;\;\;\;\;\;\;\;\;\max_{\boldsymbol{O}^{(n)}, \boldsymbol{L}^{(n)}_{\text{D}} \forall n \in \{1,\ldots,N\}} \sum_{n=1}^{(n)} R^{(n)}\\
    &\text{subject to\;\;} \forall n \in \{1,\ldots,N\}:\\
    &\cos^{-1}\left(\frac{\text{tr}\left(\boldsymbol{O}^{(n)} \left(\boldsymbol{O}^{n+1}\right)^{-1}\right)}{2}\right) \leq \Gamma_D \times T_s \label{eqn:opt0_cond0}\\
    &\left\| \left( x_D^{n+1} - x_D^{(n)}, y_D^{n+1} -  y_D^{(n)}, z_D^{n+1} - z_D^{(n)} \right) \right\| \leq v_D \times T_s \label{eqn:opt0_cond1}\\
    &x_{\min} \leq x^{(n)} \leq x_{\max} \label{eqn:opt0_cond2}\\
    &y_{\min} \leq y^{(n)} \leq y_{\max} \label{eqn:opt0_cond3}\\
    &z_{\min} \leq z^{(n)} \leq z_{\max} \label{eqn:opt0_cond4}
\end{align}
\end{subequations}

The goal, as defined in \eqref{eqn:optimization0}, is to optimize the cumulative throughput over the time span of $N T_s$. The constraints are:
\begin{itemize}
    \item \eqref{eqn:opt0_cond0} uses the matrix trace operator $\text{tr}(\cdot)$ to compute the rotation angle from the rotation matrices and restrict it to a maximum value, thereby adhering to the DRS's maximum rotational speed limit.
    \item \eqref{eqn:opt0_cond1} confirms the DRS's displacement per time step stays within its maximum translational speed.
    \item Constraints \eqref{eqn:opt0_cond2}, \eqref{eqn:opt0_cond3}, and \eqref{eqn:opt0_cond4} ensure the DRS's movement is confined within the designated 3D boundary.
\end{itemize}

The problem is divided into two subproblems: the first problem deals with deciding the trajectory of the DRS, $\boldsymbol{L^n}_{\text{D}}(t)$, by guiding it towards the optimal locations for maximizing the received power from the transmitter at $\boldsymbol{L}_{1}(t)$ to the receiver at $\boldsymbol{L}_{2}(t)$; and the second problem is concerned with determining the DRS orientation at each time step to minimize the interference from a nearby node, identified by its location $\boldsymbol{L}_{I}(t)$. The solutions are introduced below.

\section{Proposed Solution}
\label{sec:solution}

To understand the solutions, note that the path-loss formula in \eqref{eq:ris_ff_ploss} can be divided into two main components: the first depends on the locations, involving the elevation angles and distances from the RIS to each vehicle, while the second part (\(\Psi\) in \eqref{eq:ris_ff_ploss}) is influenced by the orientation, primarily depending on the azimuth angles.

The first part of the path-loss in \eqref{eq:ris_ff_ploss} is influenced only by changes in the DRS's position relative to the V2V pair. Therefore, it can be optimized independently as follows.

\subsection{Setting the trajectory of the DRS}
The optimal location is derived similarly to the approach in \cite{uav_ris_own}. Initially, the elevation angle $\theta$ for any vehicle relative to the RIS is given by:
\begin{equation}\label{eq:equal_theta}
    \theta = \tan^{-1}\left(\frac{d_\text{2d}}{h_D}\right),
\end{equation}
where $d_\text{2d}$ denotes the horizontal distance on the \(xy\)-plane between the vehicle and the DRS. The optimal $(x,y)$ location of the DRS, which minimizes \eqref{eq:ris_ff_ploss}, is identified as the midpoint between the V2X pairs. The optimal height is determined by minimizing the terms in \eqref{eq:ris_ff_ploss} excluding $\Psi$, resulting in:
\begin{equation}
    f(h_D) = \frac{d_{\text{2D}}^2+h_D^2}{\cos^6\left(\tan^{-1}\left(\frac{d_{\text{2D}}}{h_D}\right)\right)},
\end{equation}\label{eq:equal_d}
where equations \eqref{eq:equal_d} and \eqref{eq:equal_theta} simplify the function to its current form, and $d_{\text{2D}}$ is the 2D distance between the vehicle and the DRS at the midpoint. Gradient-based optimization methods such as truncated Newton conjugate-gradient (TNC) are employed to find the optimal height $h_{D, \text{opt}}$, using Scipy's implementation of TNC to optimize within the bounded space defined by \eqref{eqn:opt0_cond4}.

The optimal location is thus:
\begin{equation}\label{eq:opt_loc}
    \boldsymbol{L}^{(n)}_{\text{opt}} = \left(\frac{x^{(n)}_1+x^{(n)}_2}{2}, \frac{y^{(n)}_1+y^{(n)}_2}{2}, h_{D,\text{opt}}\right),
\end{equation}
which informs the DRS's trajectory at each time step, calculated as:
\begin{equation}\label{eq:loc_update}
    \boldsymbol{L}^{n+1} = \boldsymbol{L}^{(n)} + v_D T_s \frac{\boldsymbol{L}^{(n)}_{\text{opt}} - \boldsymbol{L}^{(n)}}{\left|\left|\boldsymbol{L}^{(n)}_{\text{opt}} - \boldsymbol{L}^{(n)}\right|\right|^2}.
\end{equation}
This strategy ensures the DRS swiftly reaches the optimal position.

\subsection{Reducing interference by controlling orientation}
To reduce interference, the orientation of the DRS, $\boldsymbol{O}^{(n)}$, is varied within the limits of its speed, $\Gamma_D$, as follows. Given the current location of the DRS, the location of the interfering node, and the receiver, then rotating the DRS with a value of $\alpha$ [rad] results in modifying the $\Psi$ term according to the following relation.
\begin{equation}\label{eq:rotate}
\begin{split}
    & \Psi_I^{(n)} (\alpha)  = \\& \frac{sinc(\frac{M\pi}{\lambda}(\sin{\theta_I}\cos{(\varphi_I + \alpha)}+\sin{\theta_r}\cos{(\varphi_r + \alpha)})d_x)}{sinc(\frac{\pi}{\lambda}(\sin{\theta_I}\cos{(\varphi_I + \alpha )}+\sin{\theta_r}\cos{(\varphi_r + \alpha)})d_x)} \\
    &\times \frac{sinc(\frac{N\pi}{\lambda}(\sin{\theta_I}\sin{(\varphi_I + \alpha)}+\sin{\theta_r}\sin{(\varphi_r + \alpha}))d_y)}{sinc(\frac{\pi}{\lambda}(\sin{\theta_I}\sin{(\varphi_I + \alpha)}+\sin{\theta_r}\sin{(\varphi_r + \alpha)})d_y)}.
    \end{split}
\end{equation}
Note that guiding this term to zero results in a large path-loss value (\eqref{eq:ris_ff_ploss}), thus cancelling the interference from the interfering location. Then we need to solve for an $\alpha < \Gamma_D \times T_s$ that will make $\Psi_I^{(n)} (\alpha)  = 0$. Since forcing any of the two numerators in \eqref{eq:rotate} to zero will effetively guide the path-loss to $\infty$, then our goal is to achieve one of the following conditions:
\begin{equation}
    C_1=\frac{M\pi}{\lambda}(\sin{\theta_I}\cos{(\varphi_I + \alpha)}+\sin{\theta_r}\cos{(\varphi_r + \alpha)})d_x = N \pi
\end{equation}
or 
\begin{equation}
    C_2=\frac{N\pi}{\lambda}(\sin{\theta_I}\sin{(\varphi_I + \alpha)}+\sin{\theta_r}\sin{(\varphi_r + \alpha}))d_y = N\pi.
\end{equation}
for any $N \in \{1,2, ...\}$
Through leveraging the trigonometric law of product, and the formula
\begin{equation}
    A \sin(x) + B \cos(x) = R \cos(x - \phi)
\end{equation}
where $R = \sqrt{A^2 + B^2}$ and $\phi = \tan^{-1}\left(\frac{A}{B}\right)$, $C_1$ and $C_2$ can be rewritten as
\begin{equation}
     C_{1,2}= D_{1,2}\sqrt{A_{1,2}^2 + B_{1,2}^2} \cos \left(\alpha-\tan^{-1}\left(\frac{A_{1,2}}{B_{1,2}}\right)\right)
\end{equation}
where 
\begin{subequations}\label{eqn:res}
\begin{align}
    &D_1 =  \frac{M\pi d_x}{\lambda}\\
    &A_1 = \sin(\theta_I) \cos(\varphi_I) + \sin(\theta_r) \cos(\varphi_I)\\
     &B_1 = -\sin(\theta_I) \sin(\varphi_I) - \sin(\theta_r) \sin(\varphi_I),
    \end{align}
\end{subequations}
and 
\begin{subequations}\label{eqn:res2}
\begin{align}
    &D_2 =  \frac{N\pi d_y}{\lambda}\\
    & A_2 = \sin(\theta_I) \sin(\varphi_I) + \sin(\theta_r) \sin(\varphi_I)\\
     &B_2 = \sin(\theta_I) \cos(\varphi_I) + \sin(\theta_r) \cos(\varphi_I).
    \end{align}
\end{subequations}
Then, to find a solution to the problem, $\alpha$ can be selected within $[-\Gamma_D, \Gamma_D]$ to satisfy any of the two constraints defined by
\begin{equation}
    \alpha_{1,2} = \cos^{-1} \left(\frac{N\pi}{D_{1,2}  \sqrt{A_{1,2}^2 + B_{1,2}^2}}\right) + \tan^{-1}\left(\frac{A_{1,2}}{B_{1,2}}\right)
\end{equation}
\section{Simulation results}
\label{sec:simres}
\subsection{Simulation setup}
The scenario assumes two parallel lanes with opposite directions and a single RSU in the middle of the area. The arrival times of vehicles entering the lanes are generated according to an exponential distribution with a rate parameter $\lambda_{\text{arrival}}$, and the number of V2V communication events starting at time step $n$ is Poisson distributed with densities $\lambda_{\text{v2v}}$, and vehicles are randomly selected. The lanes are situated parallel to the y-axis at $x=x_{\min}=0$~m and $x=x_{\max}=500$~m, and they stretch across the y axis between $y=y_{\min}=0$~m and $y=x_{\max}=5000$~m. It is assumed that the DRS can fly as low as $z_{\min} = 100$~m and $z_{\max} = 600$~m. Also, time step ($T_s$), DRS speed ($v_D$, $\Gamma_D$), and vehicle speed ($v_t$) are 0.5 s, 18 m/s, 0.1745 rad/s, and 15 m/s, respectively.

\begin{figure}[H]
\centering
\includegraphics[width=0.7\linewidth,trim={0cm, 0cm, 0cm, 0cm},clip]{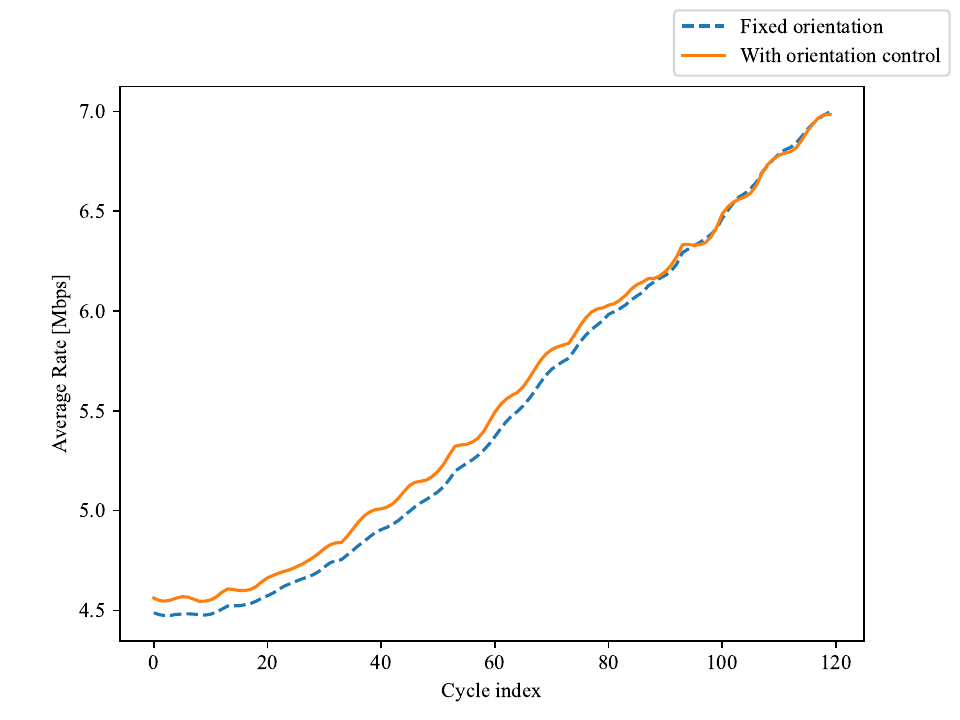}
\caption{Received throughput through the DRS at each cycle index averaged across all simulated V2V pairs. Optimizing orientation seems to slightly improve the performance since it cancels the interference.}
\label{fig:rate}
\end{figure}
\vspace{-0.5cm}
\begin{figure}[H]
\centering
\includegraphics[width=0.7\linewidth,trim={0cm, 0cm, 0cm, 0cm},clip]{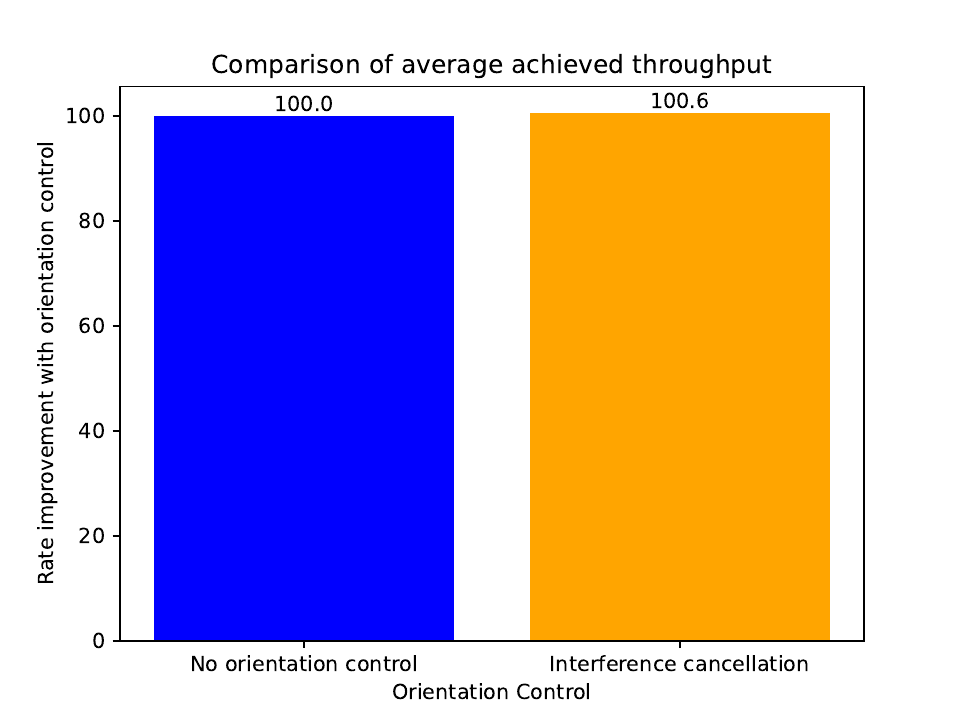}
\caption{Cancelling the interference seems to improve the rate by 0.5\%.}
\label{fig:bars}
\end{figure}
Fig.~\ref{fig:rate} plots the average rate received through the DRS across the simulation cycles with and without using orientation control. Optimizing the orientation with the described method seems to slightly improve the received rate by around 0.6\% across all simulation runs as highlighted in Fig.~\ref{fig:bars}. Although this value is small for a single simulated interference source, it nevertheless sheds light on the presented problem, and proves the formulation described in the previous section.

\section{Conclusions}
\label{sec:concl}
This paper investigated the application of DRS equipped with RIS to support V2V communications while also considering interference from a single source. A heuristic trajectory that guides the DRS to the optimal location was proposed along with an analytical solution to control the orientation of the DRS in order to cancel interference. The presented technology is simple to implement as no complex hardware is needed to be mounted on the drone. It is only needed to control the location and orientation of the drone as described. This is possible with several drones on the market. In future work, larger problem size will be considered with multiple V2V pairs and interference sources. 

The presented work has been funded by the National Science Centre in Poland within the project (no. 2021/43/B/ST7/01365) of the OPUS programme.
 
\bibliography{bibl}
\end{multicols}
\end{document}